\documentclass[sigconf]{acmart}

\usepackage[subtle]{savetrees}

\usepackage{amsmath}
\usepackage[table,dvipsnames]{xcolor}
\definecolor{Gray}{gray}{0.92}
\definecolor{yesgreen}{rgb}{0,0.6,0}
\definecolor{nored}{rgb}{0.8,0,0}
\definecolor{mycolor}{RGB}{127,8,153}
\definecolor{NavyBlue}{rgb}{0.0,0.0,0.5}

\usepackage{graphicx}
\usepackage{subcaption}
\usepackage{booktabs}
\usepackage{tabularx}
\usepackage{array}
\usepackage{makecell}
\usepackage{multirow}
\usepackage{ragged2e}

\usepackage[inline]{enumitem}
\usepackage{ifthen}
\usepackage{xspace}

\usepackage{algorithm}
\usepackage{algpseudocode}

\usepackage{listings}

\lstdefinelanguage{XML}{
  morestring=[b]",
  morestring=[s]{>}{<},
  morecomment=[s]{<?}{?>},
  stringstyle=\color{black},
  identifierstyle=\color{Blue},
  keywordstyle=\color{RoyalBlue}
}

\hypersetup{
  colorlinks=true,
  linkcolor=blue,
  citecolor=blue,
  urlcolor=blue
}

\makeatletter
\newcommand*{\etc}{%
  \@ifnextchar{.}{etc}{etc.\@\xspace}%
}
\newcommand*{\etal}{%
  \@ifnextchar{.}{et al}{et al.\@\xspace}%
}
\makeatother

\newboolean{showcomments}
\setboolean{showcomments}{false}

\ifthenelse{\boolean{showcomments}}{%
  \newcommand{\mynote}[3]{%
    \fbox{\bfseries\sffamily\scriptsize #1}%
    {\footnotesize
      $\blacktriangleright$%
      \textsf{\color{#3}#2}%
      $\blacktriangleleft$}%
  }%
}{%
  \newcommand{\mynote}[3]{}%
}

\graphicspath{{./figs/}}

\newcolumntype{C}[1]{%
  >{\centering\let\newline\\\arraybackslash\hspace{0pt}}m{#1}%
}

\AtBeginDocument{%
  }

\setcopyright{none}
\setcopyright{none}
\renewcommand\footnotetextcopyrightpermission[1]{}
\begin{document}

\title{When Do Microservices Save Energy?~Evidence from Environmental Simulation Workflows}

\author{Joshua Rowley, Abdessalam Elhabbash}
\email{joshgr704@gmail.com, a.elhabbash@lancaster.ac.uk}
\orcid{}
\affiliation{%
  \institution{School of Computing and Communications, Lancaster University, UK}
  \country{}
}

\renewcommand{\shortauthors}{Rowley and Elhabbash}

\begin{abstract}
Environmental simulation models support scenario analysis, calibration, and decision-making, but repeated execution can incur significant energy costs. Microservices offer modularity and scalability, yet their low-carbon impact remains unclear because decomposition introduces orchestration, communication, persistence, and idle-service overheads. This paper evaluates four environmental models as containerised microservice workflows, comparing monolithic execution with polling-based and event-driven orchestration. Results show that microservices increase energy consumption for smaller or tightly coupled models, where coordination overhead dominates. For a larger workflow, event-driven orchestration reduces energy use despite longer runtime, while selective downstream re-execution achieves a 41\% reduction during repeated parameter exploration.
\end{abstract}

\keywords{Energy-aware computing, Environment models, Microservices}

\maketitle

\setcopyright{none}
\settopmatter{printacmref=false}
\renewcommand\footnotetextcopyrightpermission[1]{}
\pagestyle{plain}

\vspace{-8pt}

\section{Introduction}

Environmental simulation models are increasingly used to support scenario analysis, calibration, sensitivity analysis, and environmental decision-making \cite{Holzworth2011}. These activities often require repeated execution of computational workflows under changing parameters, making execution efficiency and energy consumption important concerns for sustainable scientific computing. This is particularly relevant for environmental models, where computational systems are used to study ecological and environmental processes, yet the execution of those systems may itself contribute to energy demand.

Many environmental models are implemented as monolithic applications, often developed over time by small interdisciplinary teams. While monolithic implementations can be efficient for local execution, they typically provide limited support for modular execution, independent scaling, reuse of intermediate results, and selective recomputation \cite{Blair2019}. These limitations become more significant when models are repeatedly executed for calibration or parameter exploration. Microservice architectures offer a possible alternative by decomposing models into independently deployable workflow stages. This can improve modularity, flexibility, and reuse, while potentially enabling more energy-aware execution strategies \cite{Qianhui2026}. However, decomposition also introduces additional costs, including orchestration, serialisation, persistence, inter-service communication, and idle-service overheads. As a result, it is unclear whether microservice deployment reduces or increases energy consumption for environmental modelling workloads.

This paper investigates the energy trade-offs of deploying environmental simulation models as containerised microservice workflows. We study four environmental models with different computational and structural characteristics: Bioturbation~\cite{Harrison2022}, Fragment-MNP~\cite{Harrison2025}, NanoFASE~\cite{Harrison2021}, and UTOPIA~\cite{Domercq2025UTOPIA}. These models vary in complexity, runtime, coupling, statefulness, and decomposition potential, making them suitable for evaluating when microservice deployment may be beneficial or harmful from a low-carbon computing perspective. We compare the original monolithic implementations against two microservice orchestration strategies: a polling queue-based architecture and an event-driven stream-based architecture. The evaluation considers execution time, CPU utilisation, architectural overhead, and estimated energy consumption across repeated model executions.

The work is guided by the following research questions:

\textbf{RQ1:} How does decomposing environmental models into microservice workflows affect estimated energy consumption compared with monolithic execution?

\textbf{RQ2:} How do polling-based and event-driven orchestration strategies influence the relationship between runtime, CPU utilisation, architectural overhead, and energy consumption?

\textbf{RQ3:} Can partial re-execution and reuse of intermediate workflow results reduce energy consumption?% during repeated parameter exploration?

This paper makes the following contributions:

\begin{itemize}
    \item We design and implement containerised microservice workflow versions of four environmental models, enabling comparison with their original monolithic implementations.
    
    \item We evaluate the energy and performance trade-offs of two orchestration strategies: a polling queue-based architecture and an event-driven stream-based architecture.
    
    \item We show that microservice deployment is not inherently energy efficient: for smaller or tightly coupled models, coordination overhead can increase total energy consumption.
    
    \item We demonstrate that, for a larger environmental workflow, event-driven orchestration can reduce energy consumption compared with monolithic execution despite longer runtime.
    
    \item We show that selective re-execution of downstream workflow stages can reduce energy consumption during repeated parameter exploration, achieving a 41\% reduction in some cases.
\end{itemize}

Overall, the paper argues that low-carbon deployment of environmental models requires workload-aware architectural design. Microservices can increase energy consumption when orchestration overhead dominates, but can support lower-energy execution when applied to iterative workflows where event-driven coordination and reuse of intermediate results avoid unnecessary recomputation.

\vspace{-8pt}
\section{Background and Research Gap}

Environmental simulation models are commonly executed as scientific workflows, where tasks are connected through data and control dependencies. These workflows support scenario exploration and decision-making, often requiring repeated execution under changing parameters. Consequently, execution cost is not only a matter of runtime or scalability, but also of energy consumption, making environmental modelling an important use case for low-carbon scientific computing \cite{largeScientificModels}.

Many environmental models are deployed as monolithic applications. While monoliths can be efficient by avoiding network communication, distributed coordination, and external persistence overheads, they often tightly couple model stages, data structures, and execution logic. This limits partial execution, independent scaling, and reuse of intermediate results, particularly when parameter changes affect only downstream computations \cite{sciWorkflowsChallenges}.

Microservice architectures offer an alternative by decomposing models into independently deployable workflow stages, such as parameter initialisation, object generation, rate calculation, solving, and visualisation. Model executions can also be represented as directed acyclic graphs, allowing dependencies, execution order, and parallelism to be defined independently of the orchestration architecture \cite{sciWorkflowsFuture}. This can support modularity and selective execution, but also introduces energy-relevant overheads.

These overheads arise from API or message-broker communication, serialisation, persistence, and orchestration. They can increase both runtime and energy consumption, especially for small models where coordination costs dominate useful computation \cite{10160171, 7922500, 9095638}. Therefore, microservices should not be assumed to be inherently low-carbon; their impact depends on model structure, decomposition granularity, orchestration strategy, and opportunities for result reuse.

Orchestration strategy is particularly important. Polling-based queues can increase idle activity as workers repeatedly check for tasks, while event-driven architectures allow workers to react to message streams and may reduce unnecessary coordination overhead. We compare these strategies using Redis-backed persistence and messaging, FastAPI-based service communication, and local Docker Compose deployment \cite{9717259, 8928192, 7830692, 8890660}.

Energy-aware evaluation must consider both computation and architectural overhead. Runtime alone is an insufficient proxy for energy consumption: faster execution may consume more energy through higher CPU utilisation, while slower execution may consume less energy if it avoids unnecessary computation or reduces active resource use \cite{legler2026service, CENTOFANTI2024110371}. Accordingly, we estimate energy using a CPU-based power model that distinguishes active execution from idle and architectural overhead.

\vspace{-10pt}
\section{Methodology}

The methodology consisted of three stages: model analysis, microservice decomposition, and comparative evaluation. First, an analysis of the underlying hydrological and ecological processes was conducted to ensure that each environmental model was represented accurately when decomposed into workflow services. As these domains fall outside the typical expertise of a software engineer, documentation for each open-source model repository was reviewed alongside supplementary material to establish a conceptual understanding of the simulated processes. This was followed by direct examination of the codebases to identify data representations, mathematical equations, execution workflows, and dependencies between model components. Where complex environmental science concepts corresponded with large or tightly coupled classes and methods, generative AI tools were used to assist in summarising code structure; however, all interpretations were manually reviewed to ensure correctness.

The second stage involved decomposing the monolithic models into microservice-style workflows. Since there is no universal strategy for decomposing monolithic systems, service boundaries were guided by established software engineering principles, model semantics, coupling, computational intensity, and opportunities for reuse. Services were designed to encapsulate semantically related functionality, such as parameter initialisation, object generation, rate-constant calculation, equation solving, and visualisation. Computationally intensive and potentially parallelisable operations were also considered during decomposition, both to support scalable execution and to allow the energy impact of isolated workflow stages to be evaluated. This enabled a software-engineering-led decomposition of the models while preserving their original scientific behaviour. 
% A summary of the resulting model characteristics is shown in Table~\ref{tab:example-models}.

% \begin{table}[h]
%     \centering
%     \small
%     \begin{tabularx}{\columnwidth}{lXXXX}
%         \toprule
%         \textbf{Model} &
%         \textbf{Complexity} &
%         \textbf{Execution time} &
%         \textbf{Granularity} &
%         \textbf{Statefulness} \\
%         \midrule

%         Bioturbation &
%         Low &
%         Short&
%         Low &
%         Minimal \\
        
%         NanoFASE &
%         High &
%         Variable &
%         Low &
%         High\\
        
%         Fragment-MNP &
%         Medium &
%         Short &
%         Moderate &
%         Medium \\
        
%         UTOPIA &
%         High &
%         Long &
%         High dependencies &
%         High\\
        
%         \bottomrule
%     \end{tabularx}
%     \caption{Characteristics of case study environmental models}
%     \label{tab:example-models}
% \end{table}

The third stage evaluated the performance and energy implications of alternative deployment designs. Two containerised microservice architectures were implemented and compared against the original monolithic implementations. The architectures primarily differ in their orchestration strategy: one uses a polling queue-based model, while the other uses an event-driven approach based on pre-filtered message streams to which workers subscribe. This comparison allows the effect of coordination mechanisms on execution time, CPU utilisation, architectural overhead, and estimated energy consumption to be analysed. Details of the system design are provided in Section~\ref{sec:sys-design}.

The evaluation was designed to identify when microservice deployment increases energy consumption due to orchestration and communication overheads, and when it can reduce energy consumption by enabling modular execution and reuse of intermediate results. This is particularly important for repeated environmental modelling workloads, where parameter tuning or scenario exploration may require only downstream stages of a workflow to be recomputed.

\vspace{-12pt}
\section{System Design}
%\subsection{Architecture}
\label{sec:sys-design}

%The system design was informed by a representative workflow describing the lifecycle of a submitted model execution. This workflow was used to define the core architectural services, task dependencies, and communication mechanisms between components:

\begin{figure}[b]
    \centering
    \includegraphics[width=0.8\columnwidth]{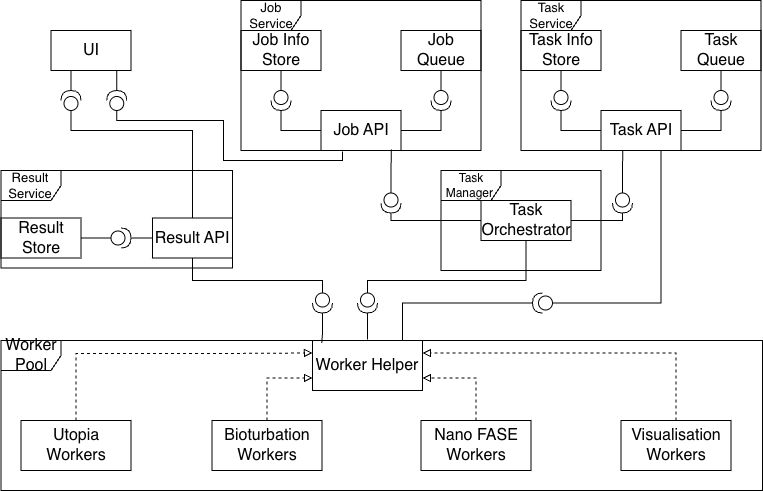}
    \caption{Ball-and-socket architecture diagram for the queue-based system.}
    \label{fig:queue_arc}
\end{figure}

% \begin{itemize}
%     \item A job is submitted via an API endpoint.
%     \item The job is stored and scheduled for processing.
%     \item The job is decomposed into one or more workflow tasks.
%     \item Worker services retrieve task parameters and any required upstream results.
%     \item Ready tasks are executed by the associated worker service.
%     \item Task outputs are persisted to a shared data store.
%     \item Persisted outputs remain available for downstream tasks, later retrieval, or partial re-execution.
% \end{itemize}
The architecture is specified using ball-and-socket diagrams to define service boundaries and inter-service communication, as shown in Fig.~\ref{fig:queue_arc}. Both microservice architectures expose lightweight API interfaces and separate model execution into jobs and tasks. The job service receives model execution requests, while the task orchestrator manages the mapping between a submitted job and the individual workflow tasks required to complete it. The result service provides access to completed outputs, for example through a user interface or external client.

Model executions are represented as directed acyclic graphs (DAGs). Each DAG defines the tasks in a model workflow, their execution order, dependency constraints, opportunities for parallelism, and result dependencies. DAGs are defined as Python objects, where tasks are enumerated and referenced before submission to the job service. This decouples workflow structure from the underlying orchestration strategy, allowing the same model workflow to be executed using either the polling queue-based architecture or the event-driven architecture without changing the job submission interface.

This design also supports energy-aware evaluation. By persisting intermediate task outputs, the system can avoid recomputing unaffected workflow stages during repeated executions. This enables partial re-execution for parameter exploration, where only downstream tasks affected by a parameter change need to be recomputed. The architecture therefore allows the energy cost of orchestration and communication to be compared against the potential energy savings from reuse and selective recomputation.

%\subsection{Decomposition}

Each model was decomposed according to model semantics, coupling structure, computational intensity, statefulness, and potential for parallel execution or reuse. The aim was to identify service boundaries that were meaningful for execution, maintainability, and energy-aware workflow evaluation.

Bioturbation was implemented as a single model service because of its low computational complexity and minimal internal coupling. Further decomposition would introduce communication and orchestration overhead without offering meaningful opportunities for parallelism or reuse. NanoFASE was decomposed into three services: initialisation, model execution, and visualisation. 
%Although NanoFASE is computationally larger, its Fortran-based implementation and state management constraints limit practical cross-service decomposition. In particular, persistent transfer of internal model state across services would require substantial refactoring, so most computation remains within a single execution service.
Fragment-MNP was decomposed into four services, reflecting a clear separation between parameter initialisation, rate-constant distribution, solving, and visualisation. %Independent rate calculations can be represented as separate tasks within the workflow DAG, enabling parallel execution where dependencies allow.
UTOPIA was decomposed into six sequential services due to its greater structural and data complexity. These services include parameter initialisation, object generation, rate-constant generation, interaction matrix construction, numerical solving, and visualisation. This decomposition provides the strongest opportunity for selective re-execution, since changes to downstream parameters can reuse earlier workflow outputs rather than recomputing the full model.

%\subsection{Implementation}

The system was implemented primarily in Python to maintain compatibility with the original model implementations. Flask was used for local interaction and monitoring, while FastAPI handled service endpoints. Inter-service communication used HTTP/1.1 requests, with Pydantic models ensuring consistent serialisation and validation of task payloads.

Docker Compose provided isolated container environments and internal service discovery. All experiments were executed locally to support controlled and reproducible comparison between monolithic and microservice deployments.

Redis acted as the shared data store and coordination mechanism. It stored task parameters, intermediate outputs, and final results, enabling downstream retrieval and partial re-execution. Redis also provided publish-subscribe and stream-based messaging for the event-driven architecture, while the queue-based architecture relied on worker polling.

Each service was implemented as a stateless worker extending a shared base class for task retrieval, dependency checking, execution, result persistence, and completion events. This ensured consistent worker behaviour across architectures.

\vspace{-10pt}
\section{Evaluation}

All experiments were conducted in a local containerised environment to ensure reproducibility and avoid variability from external cloud infrastructure. Experiments were performed on a MacBook Air with an Apple M3 processor and 16,GB of unified memory running macOS. Microservices were deployed using Docker Compose through Docker Desktop, while monolithic baselines ran as local processes on the same machine.

Each model was executed 20 times for each implementation: monolithic, queue-based microservices, and event-driven microservices. Representative parameter sets were derived from typical configurations in the original models and kept consistent across all architectural variants. Results were averaged across runs to reduce variability in execution time, resource utilisation, and communication overhead.

Evaluation was performed at both service- and system-level granularity. Service-level measurements capture individual workflow components, while system-level measurements capture end-to-end model execution. This enables comparison of localised overheads, such as inter-service communication and task orchestration, with total workflow cost.

The evaluation considered execution time, CPU utilisation, memory usage, network I/O, throughput, and estimated energy consumption. Application-level measurements were collected using Python timing and resource monitoring tools, including \texttt{time.perf\_counter()} and \texttt{resource.getrusage()}. Container-level measurements used Linux cgroup v2 interfaces, procfs, and Docker statistics to capture CPU, memory, and network behaviour across services.

Energy consumption was estimated in Joules using a CPU-based power model distinguishing active and idle CPU states. Energy in each state was calculated as power consumption in Watts multiplied by the duration spent in that state. Total energy was then computed by combining active execution energy with idle and architectural overheads, including orchestration, communication, persistence, and waiting time. These measurements enable comparison of both time-to-solution and energy-to-solution across monolithic, polling queue-based, and event-driven implementations.

\begin{figure*}[h]
\centering

\begin{subfigure}{0.4\textwidth}
    \includegraphics[width=\textwidth]{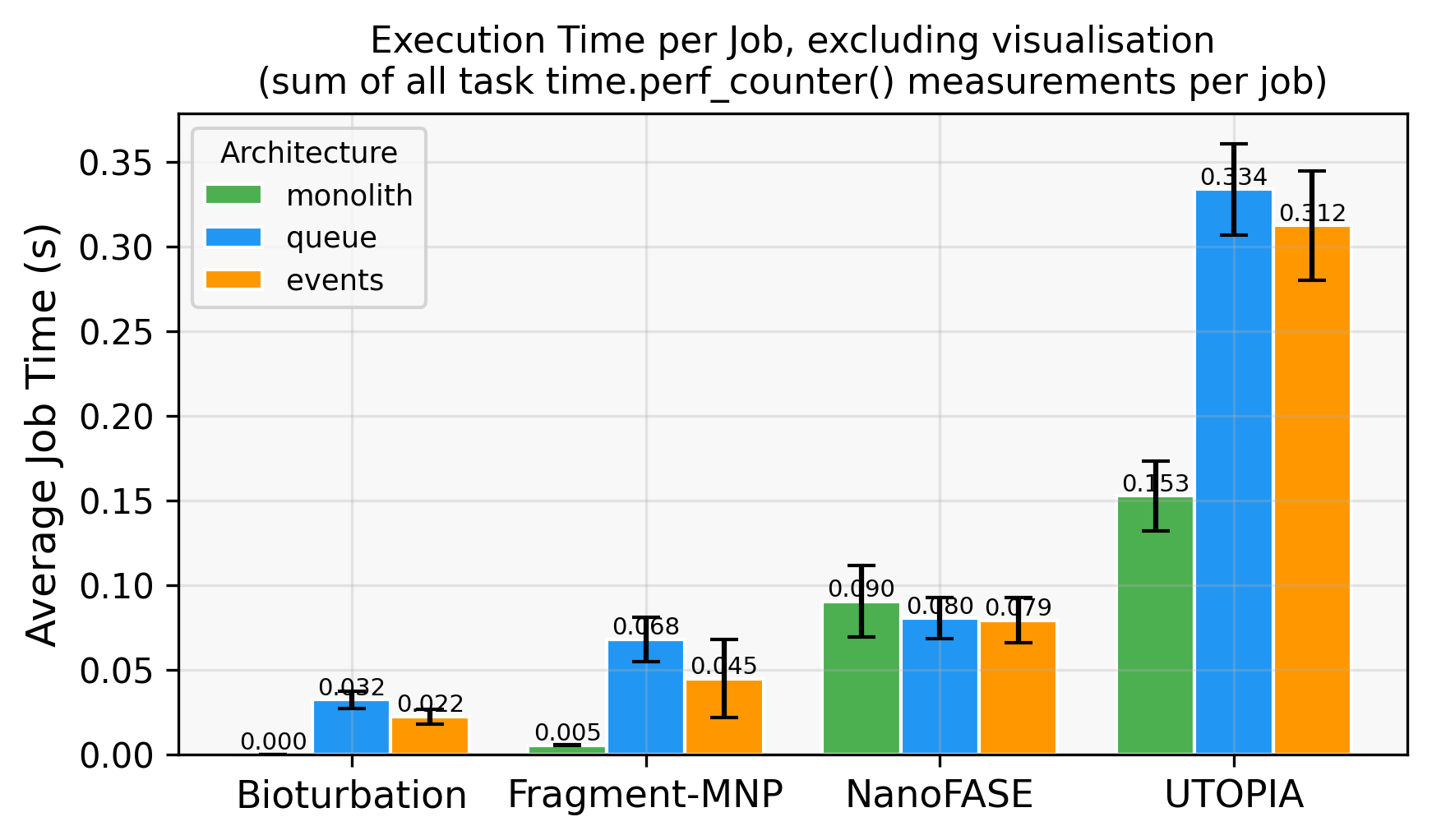}
    \caption{Average execution time per model}
    \label{fig:exec_time}
\end{subfigure}
\hfill
\begin{subfigure}{0.41\textwidth}
    \includegraphics[width=\textwidth]{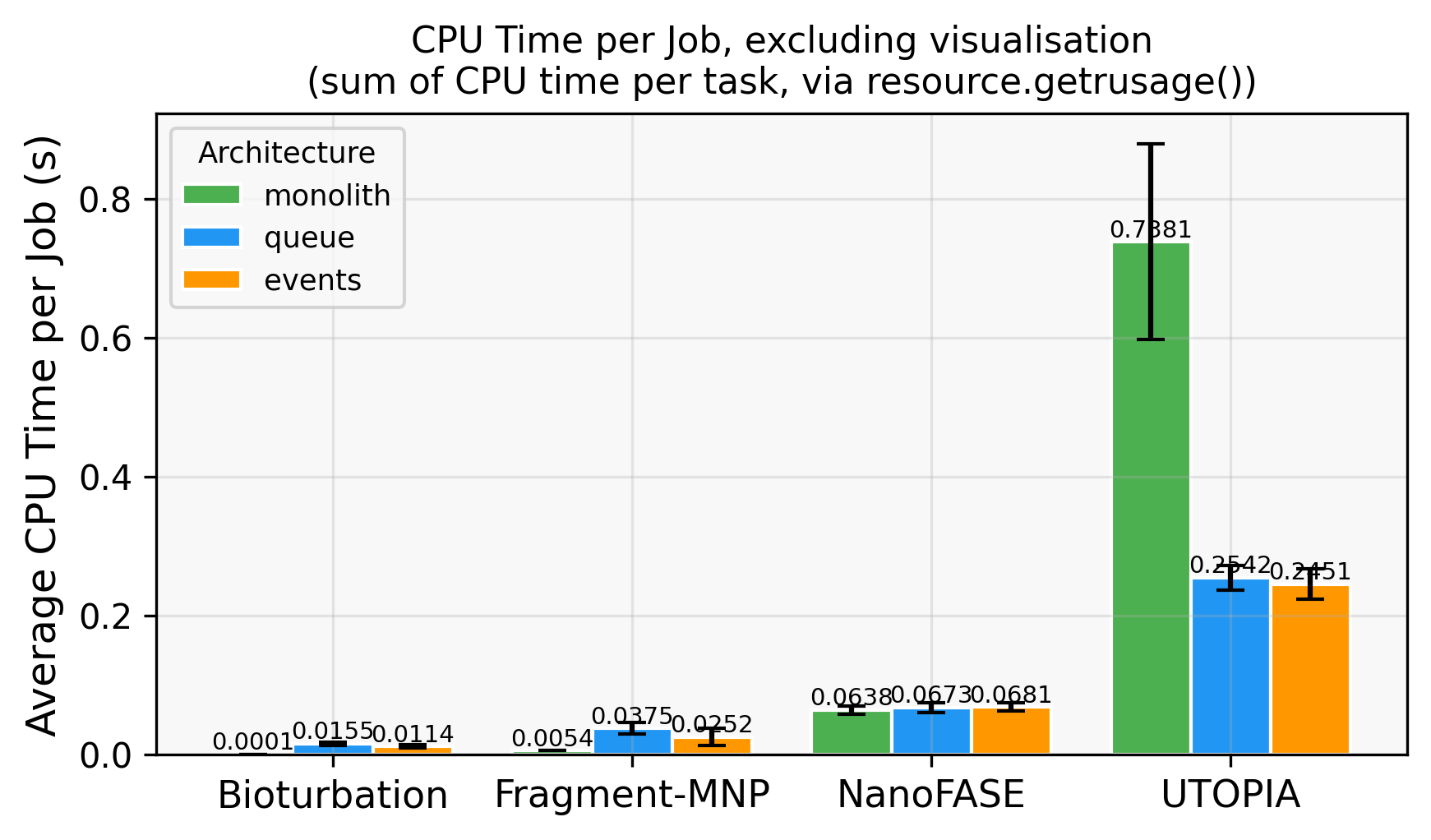}
    \caption{Average CPU time per model}
    \label{fig:cpu_time}
\end{subfigure}

\vspace{0.2cm}

\begin{subfigure}{0.43\textwidth}
    \includegraphics[width=\textwidth]{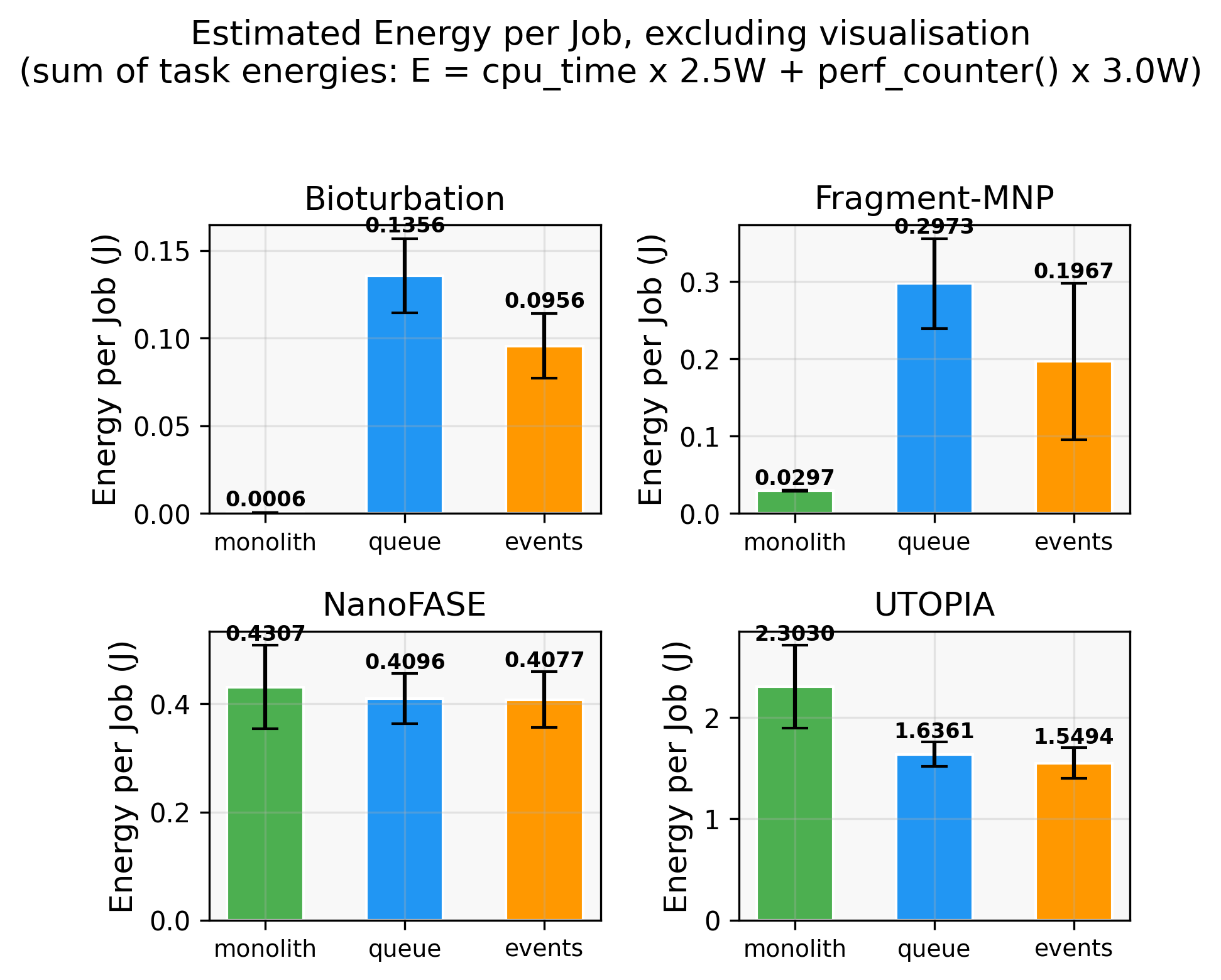}
    \caption{Energy per execution without overhead}
    \label{fig:model_energy_no_overhead}
\end{subfigure}
\hfill
\begin{subfigure}{0.43\textwidth}
    \includegraphics[width=\textwidth]{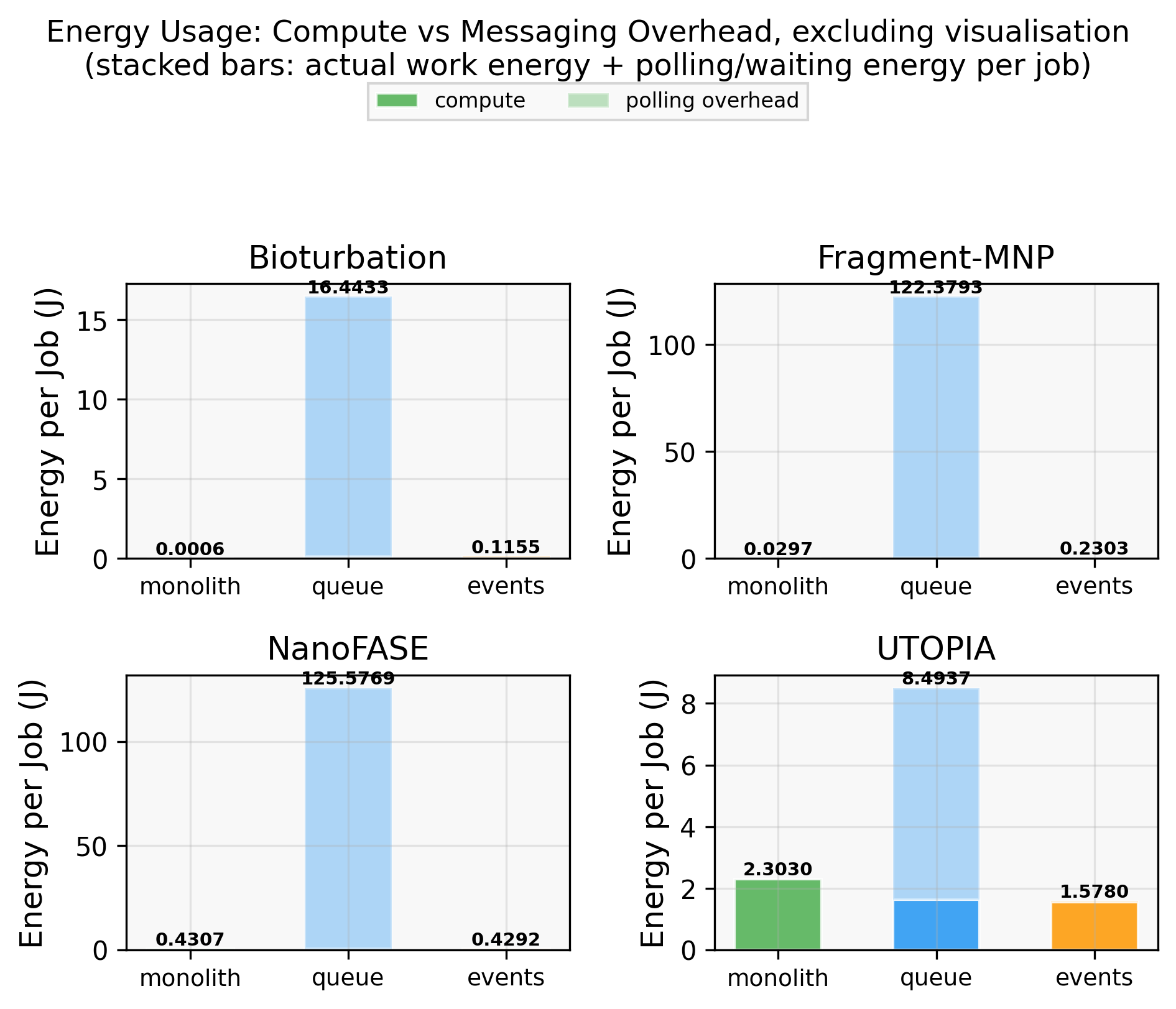}
    \caption{Energy per execution with overhead}
    \label{fig:model_energy_with_overhead}
\end{subfigure}

\vspace{0.2cm}

\begin{subfigure}{0.4\textwidth}
    \centering
    \includegraphics[width=\textwidth]{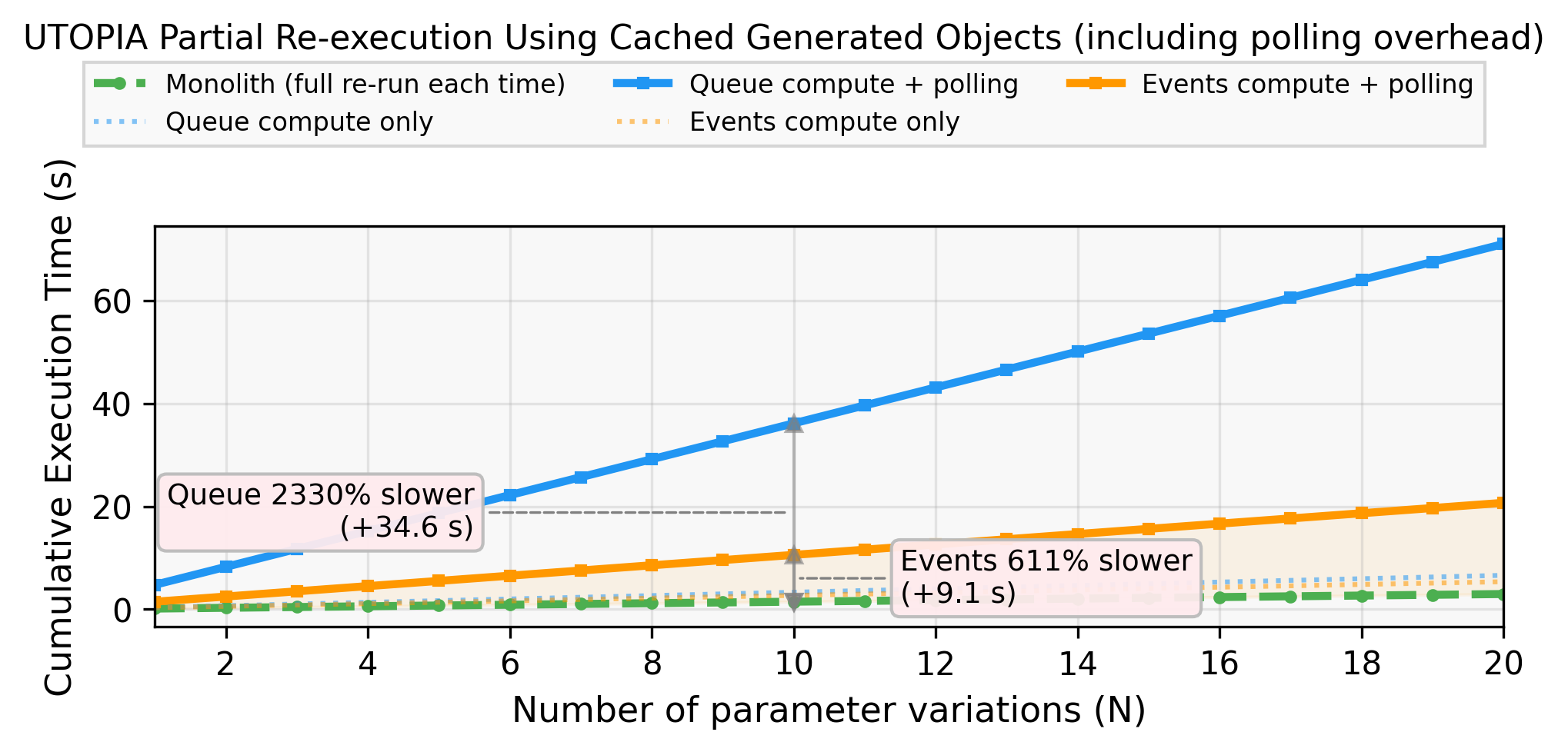}
    \caption{Effect on Time for 20 re-executions of UTOPIA (including network overhead)}
    \label{fig:utop_rerun_time}
\end{subfigure}
\hfill
\begin{subfigure}{0.4\textwidth}
    \centering
    \includegraphics[width=\textwidth]{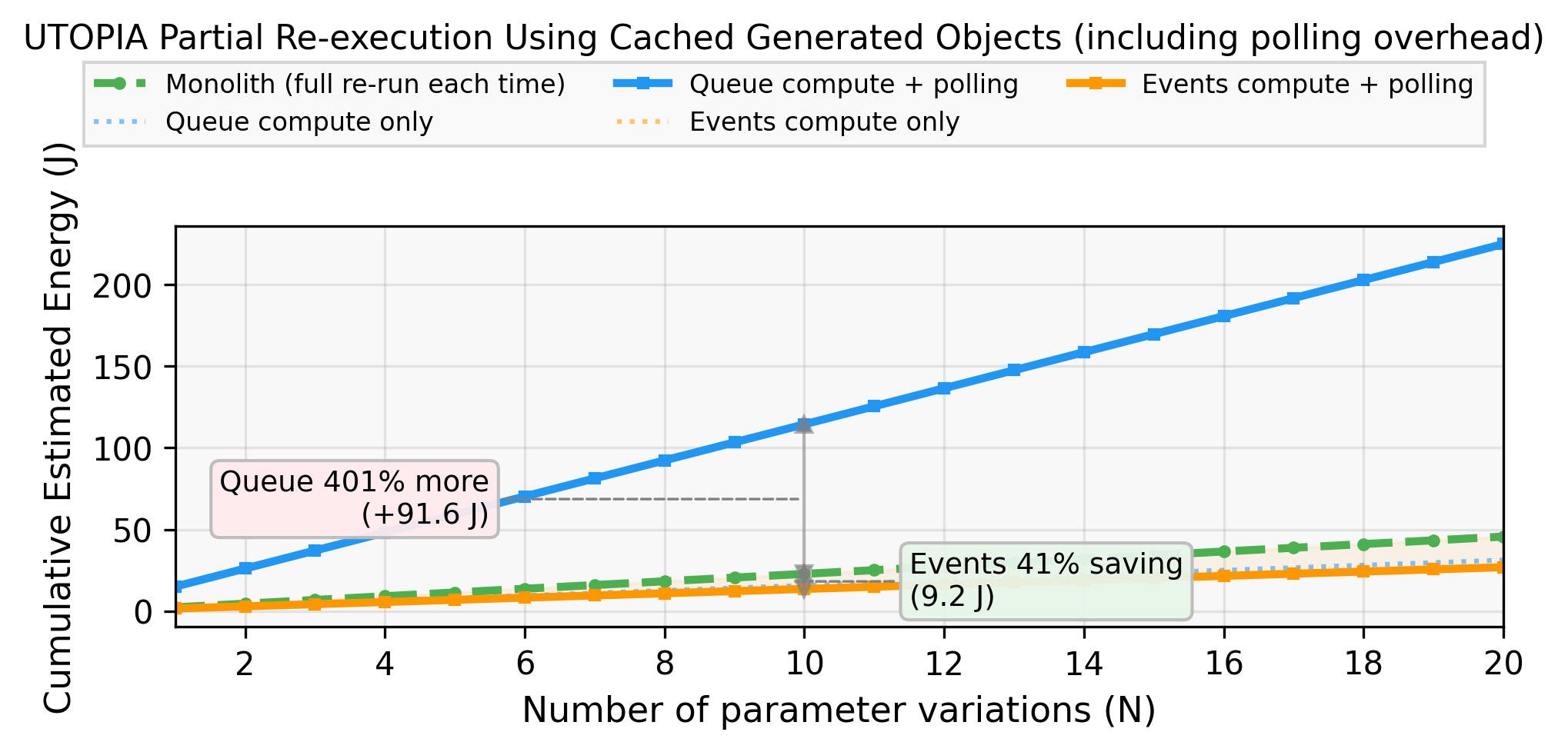}
    \caption{Effect on Energy for 20 re-executions of UTOPIA (including network overhead)}
    \label{fig:utop_rerun_energy}
\end{subfigure}

\caption{Model execution behaviour across monolithic and microservice implementations.}
\label{fig:results}
\end{figure*}

\vspace{-10pt}
\section{Results}

Microservice deployment generally increased execution time relative to the monolithic baseline, as shown in Fig.~\ref{fig:exec_time}. This was most pronounced for Bioturbation and Fragment-MNP, where limited computational work meant that orchestration and communication overhead dominated execution. CPU time in Fig.~\ref{fig:cpu_time} follows a similar trend, with both smaller models showing increased CPU usage under microservice deployment. NanoFASE showed little variation across architectures, reflecting its relatively coarse decomposition. UTOPIA was the only model to show reduced CPU time under microservice deployment, suggesting that decomposition and container-level execution changed the resource utilisation profile of the workflow.

Figures~\ref{fig:model_energy_no_overhead} and~\ref{fig:model_energy_with_overhead} compare estimated energy consumption with and without architectural overhead. For Bioturbation and Fragment-MNP, both microservice architectures consumed more energy than the monolithic baseline, indicating that coordination overhead outweighed any benefit from decomposition. NanoFASE showed similar execution energy across architectures, but total energy increased once orchestration and communication overheads were included.

UTOPIA exhibited a different pattern. Without architectural overhead, both microservice implementations consumed less energy than the monolithic baseline, as shown in Fig.~\ref{fig:model_energy_no_overhead}. When overhead was included, the event-driven architecture remained more energy efficient than the monolithic implementation, while the queue-based architecture became the most energy-consuming option, as shown in Fig.~\ref{fig:model_energy_with_overhead}. This indicates that larger workflows can benefit from decomposition, but only when orchestration overhead is sufficiently controlled.

To address RQ3, partial re-execution was evaluated using UTOPIA, as shown in Fig.~\ref{fig:utop_rerun_time} and Fig.~\ref{fig:utop_rerun_energy}. Parameters affecting downstream workflow stages were modified, allowing outputs from earlier stages such as initialisation, object generation, and rate-constant generation to be reused. Although the event-driven architecture required approximately six times longer execution than the monolithic baseline, it achieved a 41\% reduction in energy consumption after 20 partial reruns. The queue-based architecture also supported reuse, but continued to consume more time and energy due to polling overhead. These results show that selective re-execution can offset some of the energy costs of microservice deployment for larger, iterative environmental modelling workflows.

\vspace{-10pt}
\section{Discussion and Conclusions}

The results show that microservice deployment is not inherently energy efficient. Its effectiveness depends on workflow granularity, model complexity, orchestration strategy, and the opportunity to reuse3intermediate results. For tightly coupled models, such as Bioturbation and Fragment-MNP, microservice deployment increased both execution time and energy consumption. In these cases, the workload was too small to offset orchestration, serialisation, persistence, and inter-service communication costs, as shown in Fig.~\ref{fig:exec_time} and~\ref{fig:model_energy_with_overhead}. This addresses RQ1 by showing that decomposition can increase energy consumption when coordination overhead dominates computation.

The choice of orchestration strategy was also significant. The queue-based architecture incurred the highest total energy consumption in several cases, largely due to polling and idle worker activity. In contrast, the event-driven architecture reduced coordination overhead and generally remained closer to the monolithic baseline, as shown in Fig.~\ref{fig:model_energy_with_overhead}. This addresses RQ2 by demonstrating that communication and coordination patterns directly affect the energy profile of microservice workflows. Event-driven orchestration appears better suited to low-carbon workflow execution than polling-based coordination.

The UTOPIA results highlight a distinction between time-to-solution and energy-to-solution. The monolithic implementation achieved shorter execution time by making greater use of available CPU resources, as shown in Fig.~\ref{fig:cpu_time} and Fig.~\ref{fig:exec_time}. However, this higher resource utilisation corresponded with greater energy consumption. The event-driven microservice implementation consumed less energy than the monolithic baseline despite longer runtime, showing that faster execution does not necessarily imply lower energy use. For low-carbon computing, deployment decisions should therefore consider energy-to-solution alongside runtime.

Partial re-execution further shows where microservice workflows can provide energy benefits. By persisting intermediate outputs, the UTOPIA workflow avoided recomputing unaffected stages when downstream parameters changed. As shown in Fig.~\ref{fig:utop_rerun_time} and Fig.~\ref{fig:utop_rerun_energy}, this enabled selective recomputation during repeated parameter exploration, achieving a 41\% reduction in energy consumption after 20 partial reruns. This addresses RQ3 and is relevant to environmental modelling workloads, where calibration, sensitivity analysis, and scenario exploration often involve repeated executions with small parameter changes.

Overall, this paper shows that microservice architectures introduce measurable overhead and should not be treated as a universal low-carbon solution. For small models, architectural costs can outweigh any benefit from decomposition. However, for larger and more modular workflows, especially those involving repeated execution, event-driven microservices can reduce estimated energy consumption by limiting idle coordination and enabling reuse of intermediate results. These findings suggest that low-carbon environmental modelling requires workload-aware decomposition, energy-aware orchestration, and selective recomputation rather than simply migrating monolithic models to microservices.

Future work will evaluate these architectures in cloud and edge environments, where elastic scaling and real network conditions may affect energy behaviour. %Further research should explore caching, scheduling, and placement strategies to reduce orchestration overhead during repeated workflow execution.

\vspace{-4pt}
\bibliographystyle{ACM-Reference-Format}
\bibliography{bib}

\end{document}